\documentclass[12pt]{article}
\usepackage{graphicx,epsfig}
\usepackage{palatino}
\begin{document}
\vspace*{-1in}
\renewcommand{\thefootnote}{\fnsymbol{footnote}}
\begin{flushright}
TIFR/TH/08-44\\
hep-ph/\\
\end{flushright}
\vskip 65pt
\begin{center}
{\Large \boldmath\bf L$h_c$} \\
\vspace{8mm}
{\bf K. Sridhar\footnote{sridhar@theory.tifr.res.in}}\\
\vspace{10pt}
{\it Department of Theoretical Physics,\\
Tata Institute of Fundamental Research, \\ 
Homi Bhabha Road, Mumbai 400 005, India.}

\vspace{80pt}
{\bf ABSTRACT}
\end{center}
The production cross-section of $h_c$, the ${}^1P_1$ charmonium state, can be
predicted in Non-Relativistic QCD (NRQCD) using heavy-quark symmetry. We
show that at the Large Hadron Collider a large cross-section for this
resonance is predicted and it should be possible to look for 
the $h_c$ through it decay into $J/\psi +\pi$ even with the statistics that
will be achieved within a few months of run-time at the LHC.
\vspace{12pt}

\vspace{98pt}
\noindent

\vskip 10pt

\setcounter{footnote}{0}
\renewcommand{\thefootnote}{\arabic{footnote}}

\vfill
\clearpage
\setcounter{page}{1}
\pagestyle{plain}
\noindent Non-Relativistic QCD (NRQCD) \cite{bbl} is an effective theory 
obtained from QCD useful for understanding the physics of quarkonia. In 
this effective description, states of momenta much larger than the heavy 
quark mass, $m$ are excluded from the QCD Lagrangian and new interaction 
terms are added to account for this exclusion. A crucial parameter is the 
relative velocity, $v$, of the quarks bound in a quarkonium state 
in terms of which the quarkonium state is expanded into 
Fock-components. It turns out that the $Q \bar Q$ states appear in either 
colour-singlet or colour-octet configurations in this expansion where the 
colour-octet configuration evolves non-perturbatively into a physical 
colour-singlet state. The cross-section for the production of a quarkonium 
$H$ takes on the following factorised form:
\begin{equation}
   \sigma(H)\;=\;\sum_{n=\{\alpha,S,L,J\}} {F_n\over m^{d_n-4}}
       \langle{\cal O}^H_\alpha({}^{2S+1}L_J)\rangle
\label{e1}
\end{equation}
where $F_n$'s are the short-distance coefficients, calculable in a pertubation
theory in $\alpha_s$, and ${\cal O}_n$ are local operators of naive 
dimension $d_n$, describing the long-distance physics. The $Q \bar Q$ pair 
produced in the short-distance process has a separation of a scale much
smaller than $1/m$ which is pointlike on the scale of the quarkonium
wavefunction, which is of order $1/(mv)$. The non-perturbative
factor $\langle O^H_n\rangle$ is proportional to the probability for a 
pointlike $Q \bar Q$ pair in the state $n$ to form a bound state $H$. The
factorisation of the short-distance and long-distance parts of the 
cross-section guarantees the momentum-independence of the non-perturbative
terms. These can be, therefore, obtained from one experiment at a given
energy and used to compute the cross-section of the quarkonium state 
in a different experimental setting.

Before this effective theory approach was developed, 
the production of quarkonia was
sought to be understood in terms of the colour-singlet model \cite{berjon, 
br}. While at lower energies this model was seen to provide an adequate 
description of the data, it was seen \cite{jpsi} in the
phenomenology of large-$p_T$ $P$-state charmonium production at
the Tevatron \cite{cdf} that colour-octet operators are very significant. 
Processes involving $P$-state quarkonia do not have a consistent 
description in terms of colour singlet operators alone \cite{bbl2}. 
Surprisingly, when data on direct $J/\psi$ production and on $\psi'$
production from the CDF experiment at the Tevatron was analysed, it
was seen that it was necessary to include the colour-octet contributions
for phenomenological reasons \cite{brfl}, even though in the case of
the $S$-waves the octet contributions are sub-leading in $v$. With
the inclusion of the colour-octet contributions the full set of charmonium
production data from the CDF could be described albeit at the inclusion
of non-calculable long-distance matrix elements \cite{cgmp,cho}. It was
only the shape of the $p_T$-distributions and not the absolute normalisations 
that was a prediction of NRQCD. Consequently, independent tests of NRQCD
were necessary and several such proposals were made \cite{photo, hadro, brch, 
lep, upsilon, kls, bbly}. However, many of these proposals are not
for large-$p_T$ quarkonium production and while they may be of some
phenomenological interest they do not provide a rigourous test of NRQCD
because the NRQCD factorisation formula holds strictly only at large-$p_T$.
For a very comprehensive review of $J/\psi$ production at the Tevatron and
the related theory, see Ref. \cite{brambilla}.

One interesting test of NRQCD comes from the study of the polarisation of
$J/\psi$'s at large-$p_T$ \cite{polar1}. The production of 
large-$p_T$ $J/\psi$'s proceeds
primarily from the fragmentation of single gluons and the $Q \bar Q$ pair
produced in the fragmentation process inherits the transverse polarisation
of the gluon. The heavy-quark symmetry of NRQCD then comes into play in
protecting this transverse polarisation in the non-perturbative evolution
of the $Q \bar Q$ pair into a $J/\psi$. The large-$p_T$ $J/\psi$ is, therefore,
strongly transversely polarised. This is not true at even moderately low $p_T$
where the $J/\psi$ is essentially unpolarised. The $p_T$ dependence of the
polarisation is, therefore, a very good test of the theory \cite{polar2}.

The CDF experiment has measured the $p_T$-dependence of the polarisation
and they find no evidence for any transverse polarisation at large
$p_T$ \cite{polar3}. Given the success of NRQCD in explaining the
production cross-sections, this failure with respect to predicting the
polarisation is, indeed, a shock. It may well be that the successful
prediction of the production cross-sections of the various resonances
was fortuituous and that the effective theory is missing out on some
aspect of the physics of quarkonium formation. It could be that the mass
of the charm quark is not large enough to be treated in NRQCD. 
On the other hand, 
polarisation measurements are usually fraught with problems and it
may well be that the problem is elsewhere. Finally, the problem may well
have to do with the theoretical uncertainties in the prediction of 
polarisation. For example, the colour-singlet
channel predicts the polarisation of the $J/\psi$ to be longitudinal.
So any effect that could substantially increase the colour-singlet
contribution could change the full predictions of polarisation quite
drastically. To this end, a modified colour-singlet model with the 
production of $J/\psi$'s initiated by a scattering of a gluon with a Reggeized
gluon has been considered \cite{stirling} but parts of the diffractive
amplitudes involved in this calculation are not easily calculable.
A more direct approach would be to study the effect of higher-order
QCD corrections. These could substantially modify the theoretical expectations
regarding polarisation. Recent work \cite{gong} on NLO corrections to both 
the colour-singlet and colour-octet channels in the production of $J/\psi$ 
suggest that even these are not enough to understand the polarisation
data. The situation is somewhat different in the case of $\Upsilon$ 
production \cite{artoisenet} where the colour-singlet contribution, 
enhanced by NLO and a part of the NNLO corrections, seems to be able to account
for the data from Tevatron. For reviews of the current status of these
calculations and their experimental consequences, see Ref.~\cite{lansberg1,
lansberg2}.

In this situation, it is worthwhile looking for other tests of NRQCD which
successfully navigate between low-$p_T$ and 
polarisation. Such a suggestion had been made years ago in the context of
charmonium production at Tevatron \cite{self}: the production of $h_c$,
the ${}^1P_1$ charmonium state. In NRQCD, this state is produced in
the colour-singlet mode and through the production of an intermediate
octet ${}^1S_0$ state. The non-perturbative matrix element for the
transition of this octet state to the physical ${}^1P_1$ state can
be inferred from other non-perturbative parameters fixed at the 
Tevatron. This is a consequence of the heavy-quark symmetry of
NRQCD. Consequently, one can predict the rate for $h_c$ production
in NRQCD. In this letter, we investigate this prediction in the
context of the Large Hadron Collider (LHC). One channel which may 
be suitable for the detection of the $h_c$ is its decay into a 
$J/\psi +\pi$. The decay branching fraction for $h_c \rightarrow 
J/\psi + \pi$ has been estimated from spectroscopy.

It may be argued that the measurement of the other charmonium resonances
like the $J/\psi$, $\chi$'s and the $\psi'$ will already provide the
tests of the NRQCD factorisation formula. The non-perturbative parameters
have been determined at the Tevatron and the factorisation formula implies
that these are not momentum-dependent. So it should be possible to predict
the cross-sections for these resonances at the LHC and check for the
validity of NRQCD. While this is true, it must be remembered that for
several years quarkonium production has also been studied in terms
of a phenomenological model known as the semi-local duality model or
the colour-evaporation model \cite{semi}. In this model, it is assumed
that the open-charm cross-section integrated over the region between
$2 m$ and the open charm threshold should be equal to the sum of the
resonance cross-sections. The resonance cross-section is then some fraction
of the open charm cross-section integrated over this mass range. The 
fraction is unknown apriori but is fixed by comparing to the data -- it
is the analog of the non-perturbative parameter that appears in NRQCD 
computations of the cross-section. This approach is seen to provide
a reasonable description of the data from Tevatron \cite{gavai, halzen}.
However, it must be borne in mind that the separation into perturbative
and non-perturbative parts in this model is not rigourously provided by
a factorisation formula as in NRQCD and, consequently, the fractions 
(non-perturbative parameters) that are determined by fitting to Tevatron data
are not guaranteed to be energy-independent. If the energy dependence of
these parameters is large, then it will not be possible to use the
semi-local duality approach in any predictive way at the LHC. However,
it may so happen that, in actual practice, the energy dependence of
the fractions turns out to be small in which case semi-local duality
will be able to predict the resonance cross-sections at the LHC as
well as NRQCD can. But these predictions, in the semi-local duality approach
can be made for only those resonances which have been measured at the Tevatron.
It is not possible to predict the cross-section for particles which have
not been detected at the Tevatron within this model approach. The search
for $h_c$ at the LHC is, therefore, important in establishing NRQCD as the
correct theory of quarkonium production.

It is also worth emphasising that the $h_c$ had eluded experiments
for a long time and it is only recently that its existence has been
verified in $e^+ e^-$ experiments at the CLEO \cite{cleo}. The LHC
is expected to produce this resonance copiously and it may provide
a study of this resonance in various decay channels and may help
understand its properties.

At the LHC, the production of the $h_c$ proceeds through the following
partonic subprocesses:
\begin{eqnarray}
g + g &\rightarrow {}^1P_1^{\lbrack 1 \rbrack} + g, \nonumber \\
g + g &\rightarrow {}^1S_0^{\lbrack 8 \rbrack} + g, \nonumber \\
q(\bar q)+ g &\rightarrow {}^1S_0^{\lbrack 8 \rbrack } + 
                  q(\bar q), \nonumber \\
q + \bar q &\rightarrow {}^1S_0^{\lbrack 8 \rbrack } + g.
      \label{e2}
\end{eqnarray}
The large-$p_T$ hadronic production cross-section is given as
\begin{eqnarray}
&&{d\sigma \over dp_T}(p \bar p \rightarrow {}^1P_1 X)
=  \nonumber \\
&& \sum \int dy\int dx_1 x_1G_{a/p}(x_1) x_2G_{b/\bar p}(x_2)
 {4p_T \over 2x_1 -\bar x_T e^y}
 {d\hat \sigma \over d \hat t}(ab \rightarrow {}^{2S+1}L_J c) .
\label{e3}
\end{eqnarray}
In the above expression, the sum runs over all the initial partons
contributing to the subprocesses;
$G_{a/p}$ and $G_{b/\bar p}$ are the distributions of partons 
$a$ and $b$ in the hadrons with momentum fractions $x_1$ and
$x_2$, respectively. 
The expressions for the singlet and the octet subprocess cross-sections,
$d\hat\sigma/d\hat t$, are given in Refs.~\cite{gtw} and \cite{cho},
respectively. 

\begin{figure}[htb]
\begin{picture}(4,6)
\put(0,0){\epsfig{file=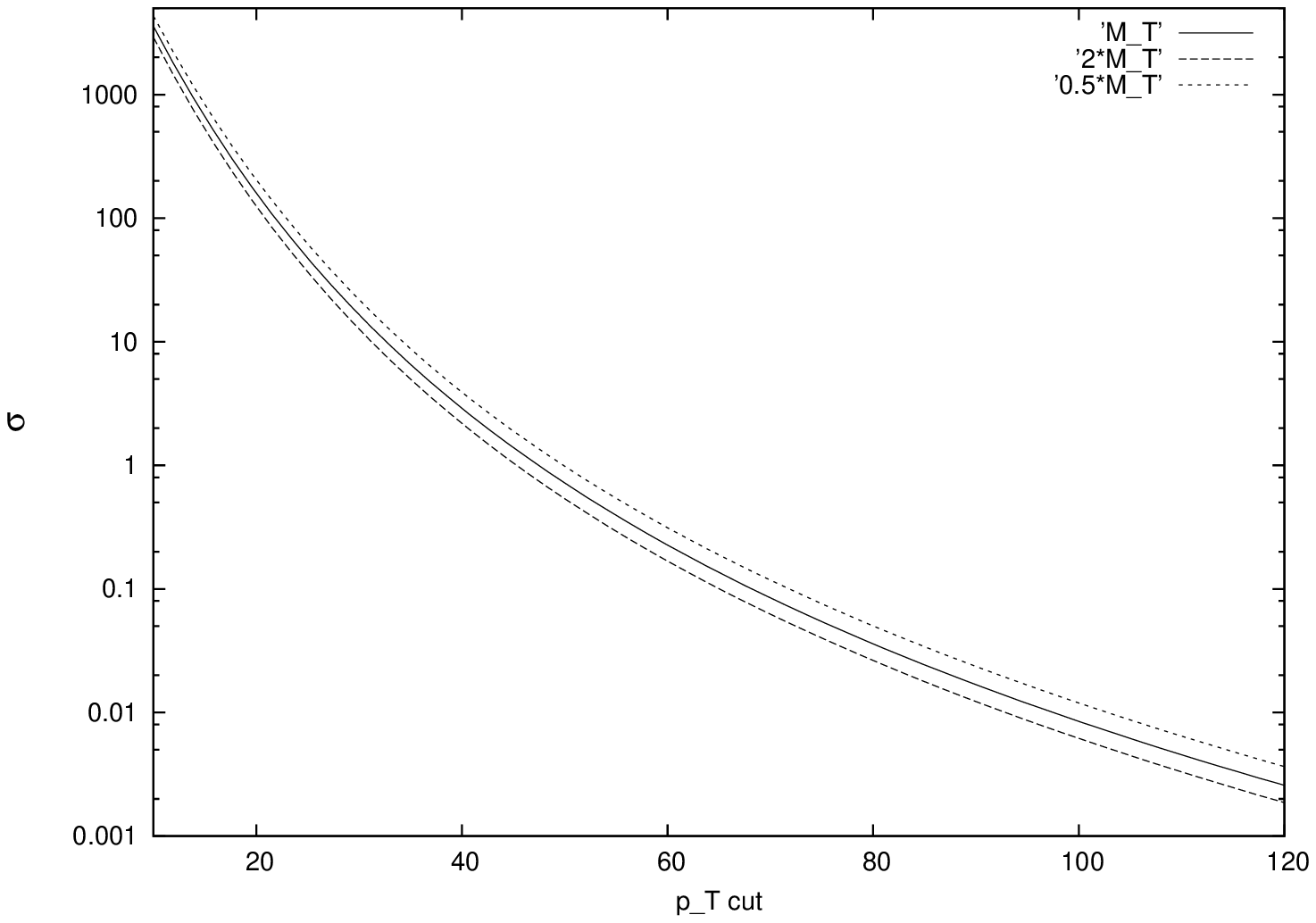, width=4in, angle=-90}}
\end{picture}
\vspace{10cm}
\caption{\it The cross-section for $h_c$ production as a function of
$p_T$ cut for different choices of QCD scale}
\protect\label{fig1}
\end{figure}
 
The ${}^1S_0^{\lbrack 8 \rbrack} \rightarrow h_c$ is mediated by
a gluon emission in a $E1$ transition. To fully determine the
production rate we need the colour-singlet matrix element for 
the ${}^1P_1$ state $\left\langle{\cal O}^{h_c}_1({}^1P_1)\right\rangle$
and the value for the colour-octet matrix element that takes the 
octet ${}^1S_0$ state to a $h_c$, $\left\langle{\cal O}^{h_c}_8({}^1S_0)
\right\rangle$.
The colour-singlet matrix element is
related to the derivative of the wavefunction of at the origin by
\begin{equation}
   \left\langle{\cal O}^{h_c}_1({}^1P_1)\right\rangle
      \;=\;{27\over2\pi}|R'(0)|^2.
\label{e6}
\end{equation}
The Tevatron data on $\chi_c$ production fixes \cite{cho} the 
colour-octet matrix element which specifies the transition of a 
${}^3S_1$ octet state into a ${}^3P_J$ state. We would expect from 
heavy-quark spin symmetry of the NRQCD Lagrangian that the 
matrix-element for ${}^1S_0^{\lbrack 8 \rbrack} \rightarrow h_c$ should be
of the same order as that for ${}^3S_1^{\lbrack 8 \rbrack} \rightarrow 
{}^3P_1$. This is because the essential difference between these
transitions comes through the magnetic quantum number so that
the corrections to this equality will be of $O(v^2) \sim$
30\%. For the derivative of the wave-function we use a similar
argument to fix it to be the same as for the $\chi_c$ states. 

\begin{figure}[htb]
\begin{picture}(4,6)
\put(0,0){\epsfig{file=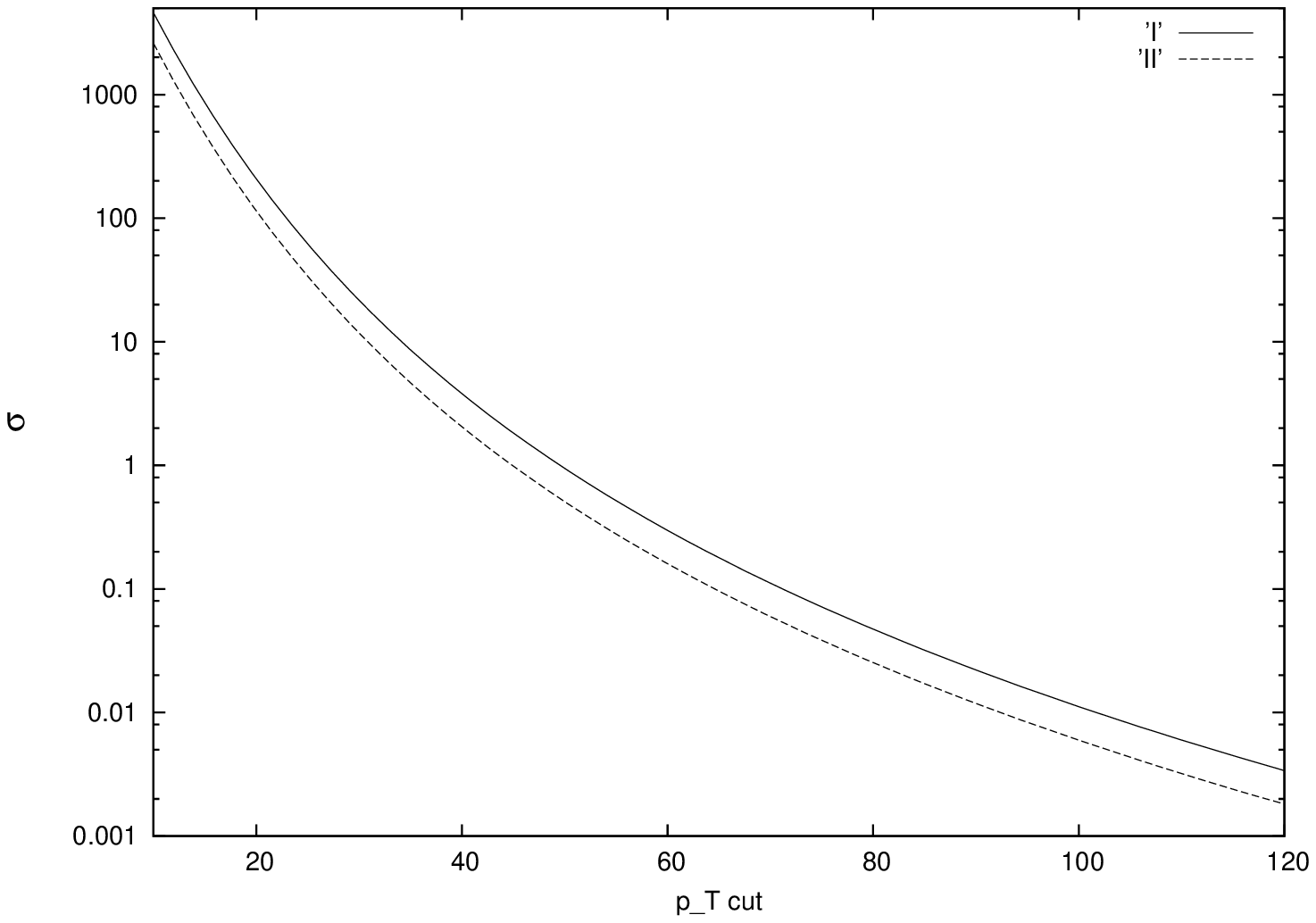, width=4in, angle=-90}}
\end{picture}
\vspace{10cm}
\caption{\it The cross-section for $h_c$ production as a function of
$p_T$ cut for the range of allowed values of the octet matrix element}
\protect\label{fig2}
\end{figure}

With these inputs, we have computed the cross-section for $h_c$ production
in $pp$ collisions at the LHC ($\sqrt{s}=14$~TeV). We have computed the
cross-section integrated over $p_T$ with a lower $p_T$ cut. In Fig. 1,
we present the results for the $p_T$-integrated cross-section as a
funtion of the $p_T$-cut for three different choices of the QCD scale:
$Q = M_T/2,\ M_T \ {\rm and}\  2M_T$. We have used the CTEQ 4M parton densities
\cite{cteq}. The cross-section has been folded in with the branching ratio 
of the ${}^1P_1$ state into $J/\psi+\pi$ and the $J/\psi \rightarrow l^+ l^-$,
where $l=e \ {\rm or}\ \mu$. We have integrated over the rapidity interval 
$-2.0 \le y \le 2.0$. For the singlet matrix element, we use the
value extracted from $\chi_c$ decays, which is 
$\left\langle{\cal O}^{h_c}_1({}^1P_1)\right\rangle=0.32$ 
\cite{mangano} and for the octet matrix element we have
$\left\langle{\cal O}^{{}^1P1}_8({}^1S_0)\right\rangle=0.0098$ 
\cite{cho}. With these inputs, we find that the cross-section for 
$h_c$ production (folded in with the decay fraction into a 
$J/\psi$ and $\pi$, which we take to be 0.5\% \cite{onep} and a 6\% leptonic
decay branching fraction of the $J/\psi$) is large enough to have a 
substantial number of events with the statistics that will be acquired
in the first few months of LHC running. Varying the QCD scale between
the largest and the smallest values that it can take results in a 
variation in the cross-section which is about a factor of 2.
While the results for the cross-section
for $h_c$ production in Fig. 1 show the variation with respect to QCD scale
inputs, in Fig.2 we display the uncertainty in the cross-section coming
from varying the value of the octet matrix element. We expect a 30\% 
variation about the central value of 0.0098 for the octet matrix element.
The two curves in Fig. 2 correspond to the upper and lower values that
the octet matrix element can take. In Fig. 2 the QCD scale is taken to be 
$M_T$. The variation in the cross-section due to the change in the octet
matrix element is about 60\%.

\begin{figure}[htb]
\begin{picture}(4,6)
\put(0,0){\epsfig{file=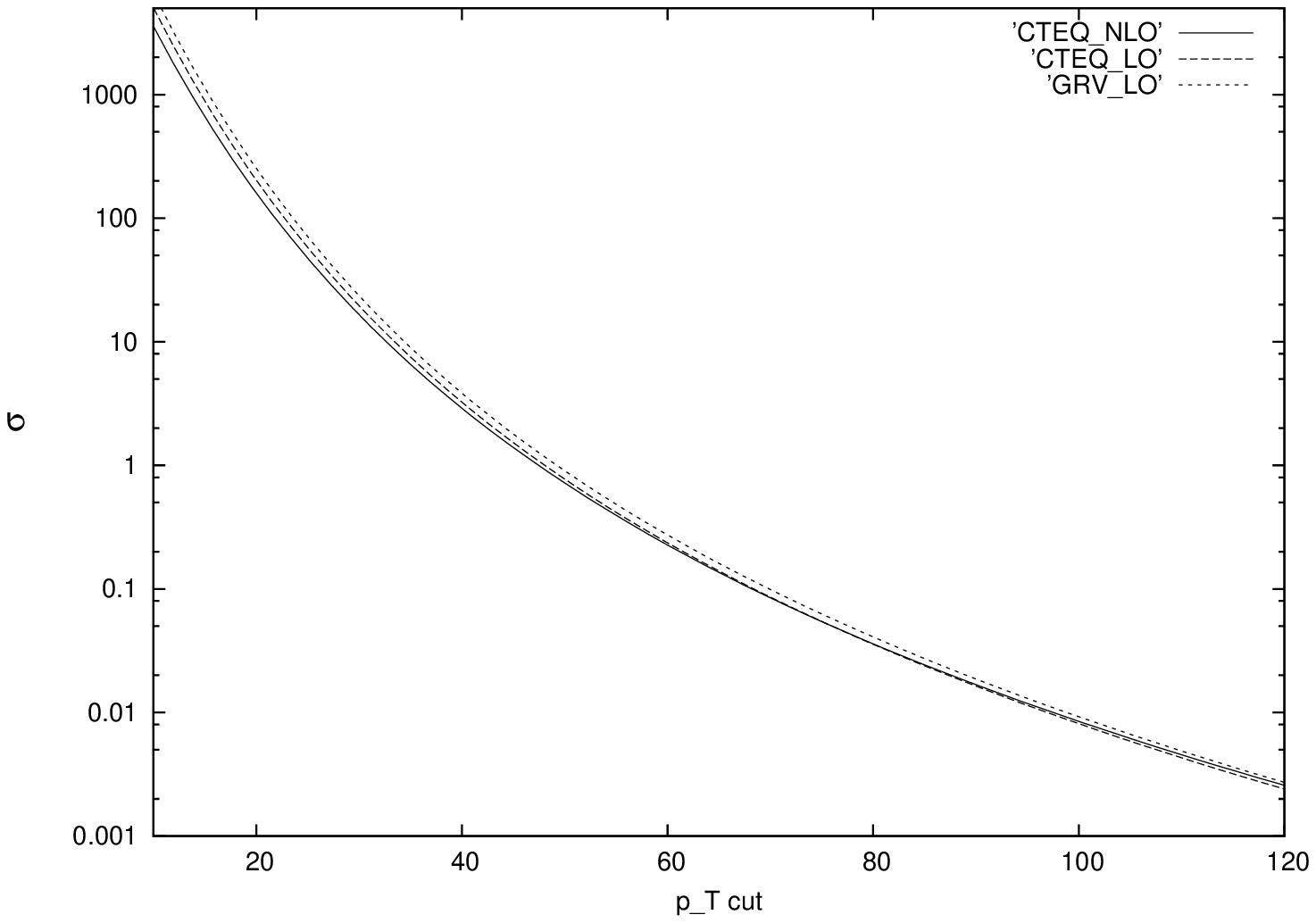, width=4in, angle=-90}}
\end{picture}
\vspace{10cm}
\caption{\it The cross-section for $h_c$ production as a function of
$p_T$ cut for different parton distribution sets}
\protect\label{fig3}
\end{figure}
 
In Fig. 3, we show the $p_T$-integrated cross-section
choosing different parton density sets. In addition to the CTEQ 4M
densities used earlier, we use the LO CTEQ \cite{cteq} and GRV densities 
\cite{grv}.
It is only at low values of $p_T$ that a sizeable change in the cross-section
due to the variation of the parton density inputs can be seen and even at a
$p_T$ value of 10 GeV the variation is not more than about 30\%.
The decay branching fraction of $h_c$ into a $J/\psi+\pi$ could
be as large as 1\% \cite{onep}, and if we use this instead of the 0.5\% 
used in the above calcuations we could have a production cross-section
which is twice as large.

The $p_T$ distribution $Bd\sigma/dp_T$ is shown in Fig. 4.  We have
plotted the octet and the singlet contributions separately. We find
that, over a whole range of large $p_T$, the singlet contribution
is negligible and that the $h_c$ is produced almost exclusively from the 
colour-octet channel.

\begin{figure}[htb]
\begin{picture}(4,6)
\put(0,0){\epsfig{file=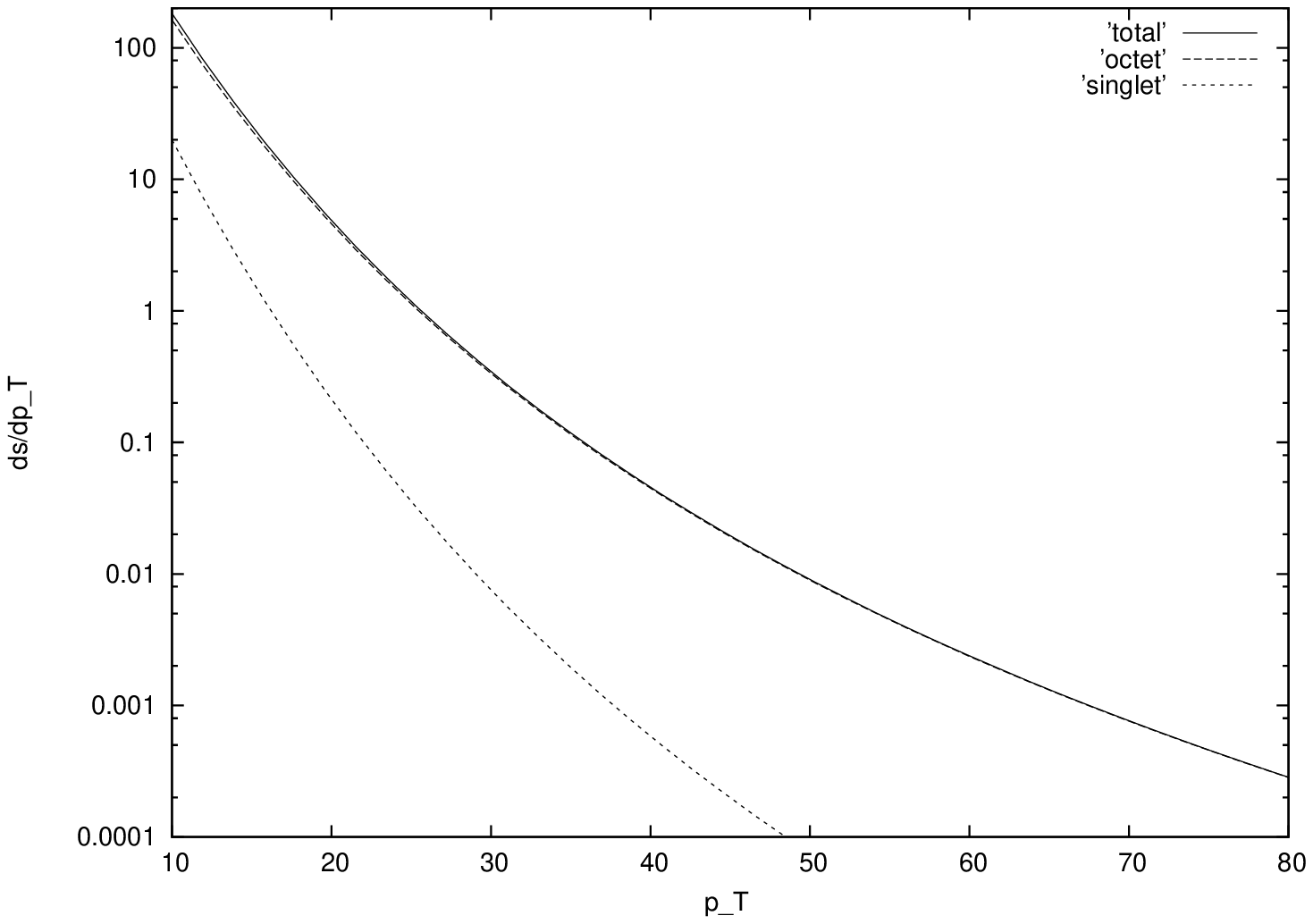, width=4in, angle=-90}}
\end{picture}
\vspace{10cm}
\caption{\it The $p_T$ distributions for $h_c$ production }
\protect\label{fig4}
\end{figure}

We would like to conclude by making the following points:
\begin{itemize} 
\item
In spite of the success that we have had in understanding charmonium
production at the Tevatron using NRQCD, we still need to have independent
tests of this effective theory because the colour-octet
parameters, and consequently the normalisations of the cross-sections
of the various charmonium resonances, are not given by the theory but only
fixed by fitting to the data.

\item
Polarisation predictions for $J/\psi$ and $\psi'$ at large=$p_T$, 
considered to be good tests of NRQCD, disagree violently with what is
measured by the CDF experiment at the Tevatron.

\item
The production of $h_c$ in NRQCD is a good test of the theory because: i)
it is a prediction for large-$p_T$ production where NRQCD factorisation
is expected to hold, ii) the cross-section can be predicted because the
relevant colour-octet parameter can be inferred from octet parameters
measured in $\chi$ production at the Tevatron and using heavy-quark 
symmetry and, iii) the cross-section is very large at the LHC and
should lead to easy detection of the resonance. Moreover, the cross-section
measurement is much simpler than measuring the polarisation of the
charmonium state.

\item
Such a prediction for the cross-section of $h_c$ production cannot
be made in the alternative approach to quarkonium production, viz., the
semi-local duality model \cite{semi, gavai, halzen}. If the prediction
of NRQCD is verified it will certainly establish it as the correct
approach to quarkonium physics.

\end{itemize} 

In conclusion, even with the statistics accumulated with a few months
of LHC running the charmonium resonance, $h_c$, can not only be detected
but its properties can be studied in detail. We have presented predictions
of NRQCD for the production cross-section of the $h_c$ and so the
study of this state at the LHC will help test NRQCD independently and
provide us more understanding of the physics of quarkonium formation.


\clearpage
\begin{thebibliography}{999}

\bibitem{bbl} G.~T.~Bodwin, E.~Braaten and G.~P.~Lepage,
  Phys.\ Rev.\  D {\bf 51} (1995) 1125
  [Erratum-ibid.\  D {\bf 55} (1997) 5853]
  [arXiv:hep-ph/9407339].

\bibitem{berjon} E.~L.~Berger and D.~L.~Jones,
  Phys.\ Rev.\  D {\bf 23} (1981) 1521.

\bibitem{br} R.~Baier and R.~Ruckl,
  Z.\ Phys.\  C {\bf 19} (1983) 251.

\bibitem{jpsi} E.~Braaten, M.~A.~Doncheski, S.~Fleming and M.~L.~Mangano,
  Phys.\ Lett.\  B {\bf 333} (1994) 548
  [arXiv:hep-ph/9405407]; D.~P.~Roy and K.~Sridhar,
  Phys.\ Lett.\  B {\bf 339} (1994) 141
  [arXiv:hep-ph/9406386]; M.~Cacciari and M.~Greco,
  Phys.\ Rev.\ Lett.\  {\bf 73} (1994) 1586
  [arXiv:hep-ph/9405241].

\bibitem{cdf} F.~Abe {\it et al.}  [CDF Collaboration],
  Phys.\ Rev.\ Lett.\  {\bf 69} (1992) 3704; 
  Phys.\ Rev.\ Lett.\  {\bf 79} (1997) 572; 
  Phys.\ Rev.\ Lett.\  {\bf 79} (1997) 578; D.~E.~Acosta {\it et al.}  [CDF Collaboration],
  Phys.\ Rev.\  D {\bf 71} (2005) 032001
  [arXiv:hep-ex/0412071].

\bibitem{bbl2} G.~T.~Bodwin, E.~Braaten and G.~P.~Lepage,
  Phys.\ Rev.\  D {\bf 46} (1992) 1914
  [arXiv:hep-lat/9205006].

\bibitem{brfl} E.~Braaten and S.~Fleming,
  Phys.\ Rev.\ Lett.\  {\bf 74} (1995) 3327
  [arXiv:hep-ph/9411365].

\bibitem{cgmp} M.~Cacciari, M.~Greco, M.~L.~Mangano and A.~Petrelli,
  Phys.\ Lett.\  B {\bf 356} (1995) 553
  [arXiv:hep-ph/9505379].

\bibitem{cho} P.~L.~Cho and A.~K.~Leibovich,
  Phys.\ Rev.\  D {\bf 53} (1996) 150
  [arXiv:hep-ph/9505329]; Phys.\ Rev.\  D {\bf 53} (1996) 6203
  [arXiv:hep-ph/9511315].

\bibitem{photo} M.~Cacciari and M.~Kramer,
  Phys.\ Rev.\ Lett.\  {\bf 76} (1996) 4128
  [arXiv:hep-ph/9601276]; J.~Amundson, S.~Fleming and I.~Maksymyk,
  Phys.\ Rev.\  D {\bf 56} (1997) 5844
  [arXiv:hep-ph/9601298].

\bibitem{hadro} S.~Gupta and K.~Sridhar,
Phys.\ Rev.\  D {\bf 54} (1996) 5545
  [arXiv:hep-ph/9601349]; Phys.\ Rev.\  D {\bf 55} (1997) 2650
  [arXiv:hep-ph/9608433]; M.~Beneke and I.~Z.~Rothstein,
  Phys.\ Rev.\  D {\bf 54} (1996) 2005
  [Erratum-ibid.\  D {\bf 54} (1996) 7082]
  [arXiv:hep-ph/9603400]; W.~K.~Tang and M.~Vanttinen,
  Phys.\ Rev.\  D {\bf 54} (1996) 4349
  [arXiv:hep-ph/9603266].

\bibitem{brch} E.~Braaten and Y.~Q.~Chen,
  Phys.\ Rev.\ Lett.\  {\bf 76} (1996) 730
  [arXiv:hep-ph/9508373].

\bibitem{lep} 
  K.~m.~Cheung, W.~Y.~Keung and T.~C.~Yuan,
  Phys.\ Rev.\ Lett.\  {\bf 76} (1996) 877
  [arXiv:hep-ph/9509308]; P.~L.~Cho,
  Phys.\ Lett.\  B {\bf 368} (1996) 171
  [arXiv:hep-ph/9509355].

\bibitem{upsilon} K.~m.~Cheung, W.~Y.~Keung and T.~C.~Yuan,
  Phys.\ Rev.\  D {\bf 54} (1996) 929
  [arXiv:hep-ph/9602423].

\bibitem{kls} P.~Ko, J.~Lee and H.~S.~Song,
  Phys.\ Rev.\  D {\bf 53} (1996) 1409
  [arXiv:hep-ph/9510202].
 
\bibitem{bbly} G.~T.~Bodwin, E.~Braaten, T.~C.~Yuan and G.~P.~Lepage,
  Phys.\ Rev.\  D {\bf 46} (1992) 3703
  [arXiv:hep-ph/9208254].

\bibitem{brambilla}
  N.~Brambilla {\it et al.}  [Quarkonium Working Group],
  arXiv:hep-ph/0412158.

\bibitem{polar1} P.~L.~Cho and M.~B.~Wise,
  Phys.\ Lett.\  B {\bf 346} (1995) 129
  [arXiv:hep-ph/9411303].

\bibitem{polar2} M.~Beneke and M.~Kramer,
  Phys.\ Rev.\  D {\bf 55} (1997) 5269
  [arXiv:hep-ph/9611218].

\bibitem{polar3} A.~A.~Affolder {\it et al.}  [CDF Collaboration],
  Phys.\ Rev.\ Lett.\  {\bf 85} (2000) 2886
  [arXiv:hep-ex/0004027]; A.~Abulencia {\it et al.}  [CDF Collaboration],
  Phys.\ Rev.\ Lett.\  {\bf 99} (2007) 132001
  [arXiv:0704.0638 [hep-ex]].

\bibitem{stirling}
  V.~A.~Khoze, A.~D.~Martin, M.~G.~Ryskin and W.~J.~Stirling,
  Eur.\ Phys.\ J.\  C {\bf 39} (2005) 163
  [arXiv:hep-ph/0410020].

\bibitem{gong}
  B.~Gong, X.~Q.~Li and J.~X.~Wang,
  arXiv:0805.4751 [hep-ph].

\bibitem{artoisenet}
  P.~Artoisenet, J.~Campbell, J.~P.~Lansberg, F.~Maltoni and F.~Tramontano,
  Phys.\ Rev.\ Lett.\  {\bf 101} (2008) 152001
  [arXiv:0806.3282 [hep-ph]].

\bibitem{lansberg1}
  J.~P.~Lansberg {\it et al.},
  AIP Conf.\ Proc.\  {\bf 1038} (2008) 15
  [arXiv:0807.3666 [hep-ph]].

\bibitem{lansberg2}
  J.~P.~Lansberg,
  arXiv:0811.4005 [hep-ph].
 
\bibitem{self} K.~Sridhar,
  Phys.\ Rev.\ Lett.\  {\bf 77} (1996) 4880
  [arXiv:hep-ph/9609285].

\bibitem{semi} H.~Fritzsch,
  Phys.\ Lett.\  B {\bf 67} (1977) 217.

\bibitem{gavai} R.~Gavai, D.~Kharzeev, H.~Satz, G.~A.~Schuler, K.~Sridhar and R.~Vogt,
  Int.\ J.\ Mod.\ Phys.\  A {\bf 10} (1995) 3043
  [arXiv:hep-ph/9502270].

\bibitem{halzen} J.~F.~Amundson, O.~J.~P.~Eboli, E.~M.~Gregores and F.~Halzen,
  Phys.\ Lett.\  B {\bf 390} (1997) 323
  [arXiv:hep-ph/9605295].

\bibitem{cleo} J.~L.~Rosner {\it et al.}  [CLEO Collaboration],
  Phys.\ Rev.\ Lett.\  {\bf 95} (2005) 102003
  [arXiv:hep-ex/0505073]; P.~Rubin {\it et al.}  [CLEO Collaboration],
  Phys.\ Rev.\  D {\bf 72} (2005) 092004
  [arXiv:hep-ex/0508037].

\bibitem{gtw} R.~Gastmans, W.~Troost and T.~T.~Wu,
  Nucl.\ Phys.\  B {\bf 291} (1987) 731.

\bibitem{cteq} H.~L.~Lai {\it et al.}  [CTEQ Collaboration],
  Eur.\ Phys.\ J.\  C {\bf 12} (2000) 375
  [arXiv:hep-ph/9903282].

\bibitem{mangano} M.~L.~Mangano and A.~Petrelli,
  Phys.\ Lett.\  B {\bf 352} (1995) 445
  [arXiv:hep-ph/9503465].

\bibitem{onep} Y.~P.~Kuang, S.~F.~Tuan and T.~M.~Yan,
  Phys.\ Rev.\  D {\bf 37} (1988) 1210.

\bibitem{grv} M.~Gluck, E.~Reya and A.~Vogt,
  Eur.\ Phys.\ J.\  C {\bf 5} (1998) 461
  [arXiv:hep-ph/9806404].

\end{thebibliography}
\end{document}